\documentclass[twocolumn,preprintnumbers,prd,superscriptaddress,nofootinbib,floatfix,showpacs,showkeys]{revtex4-1}

\usepackage{amsmath}
\usepackage{amsfonts}
\usepackage{amssymb}
\usepackage{graphicx}
\usepackage{color}
\usepackage{hyperref}
\usepackage{booktabs}
\usepackage{cleveref}
\usepackage{mathtools}
\usepackage{subfigure}
\usepackage{enumerate}

\definecolor{amaranth}{rgb}{0.9, 0.17, 0.31}
\definecolor{palatinateblue}{rgb}{0.15, 0.23, 0.89}
\definecolor{darkgreen}{rgb}{0, 0.5, 0}

\hypersetup{linktoc=all,
    colorlinks, linkcolor=palatinateblue,
    citecolor=darkgreen, urlcolor=amaranth
    }

\Crefname{equation}{Eq.}{Eqs.}
\Crefname{figure}{Fig.}{Figs.}
\Crefname{section}{Sec.}{Secs.}

\begin{document}

\title{Healing the cosmological constant problem during inflation through a unified \emph{quasi-quintessence} matter field}

\author{Rocco D'Agostino}
\email{rocco.dagostino@unina.it}
\affiliation{Scuola Superiore Meridionale, Largo S. Marcellino 10, 80138 Napoli, Italy.}
\affiliation{Istituto Nazionale di Fisica Nucleare (INFN), Sez. di Napoli, Via Cinthia 9, 80126 Napoli, Italy.}

\author{Orlando Luongo}
\email{orlando.luongo@unicam.it}
\affiliation{Dipartimento di Matematica, Universit\`a di Pisa, Largo B. Pontecorvo 5, 56127 Pisa, Italy.}
\affiliation{Universit\`a di Camerino, Divisione di Fisica, Via Madonna delle carceri 9, 62032 Camerino, Italy.}
\affiliation{NNLOT, Al-Farabi Kazakh National University, Al-Farabi av. 71, 050040 Almaty, Kazakhstan.}

\author{Marco Muccino}
\email{marco.muccino@lnf.infn.it}
\affiliation{NNLOT, Al-Farabi Kazakh National University, Al-Farabi av. 71, 050040 Almaty, Kazakhstan.}

\begin{abstract}
We heal the cosmological constant problem by means of a \emph{cancellation mechanism} that adopts a phase transition during which quantum fluctuations are eliminated. To this purpose, we propose that a generalized scalar (dark) matter field with a non-vanishing pressure term can remove the vacuum energy contribution, if its corresponding thermodynamics is written in terms of a \emph{quasi-quintessence} representation. In such a  picture, pressure differs from quintessence as it shows a zero kinetic contribution. Using this field, we investigate a metastable transition phase, in which the universe naturally passes through an inflationary phase. To reach this target, we single out a double exponential potential, describing the metastable inflationary dynamics by considering  suitable boundaries and thermodynamic conditions. We analyze stability investigating saddle, stable and unstable points and we thus predict a chaotic inflation that mimics the Starobinsky exponential potential. Consequently, the role of the proposed dark matter field is investigated throughout the overall universe evolution. To do so, we provide a physical explanation on unifying the dark sector with inflation by healing the cosmological constant problem. 
\end{abstract}


\maketitle

%%%%%%%%%%%%%%%%%%%%%%%%%%%%%%%%%%%%%%%%%%%%%%%%%%%%%%%%%%%%%%%%%%%%%%%%%%%%%%%%%%%%%%%%%%%%%%%%%%%%%%%%%%%%%%%%%%%%%%%%%%%%%%%%%%%%%%%%%%%%%%%%%%%%%%%%%%%%%%%%%%

\section{Introduction}
\label{sezione1}

Unified dark energy models aim at describing dark energy and dark
matter by means of a single fluid \cite{Scherrer04,Cardone04,Bertacca11,Brandenberger21,Anton-Schmidt,logotropic}. The universe's dynamics in the standard cosmological puzzle is characterized with pressureless dark matter and a dark energy contributions, exhibiting negative pressure\footnote{For instance, within the Chaplygin and modified Chaplygin gas models \cite{Kamenshchik01,Bento02,Sandvik04}, baryons contribute to visible matter, whereas the dark sector emerges from the Chaplygin non-vanishing pressure.}\cite{Carroll01,Padmanabhan03,Copeland06,review}. The  dark fluid \cite{Beca07,Luongo14,Mishra21} fully degenerates with the $\Lambda$CDM paradigm, so that one can naively consider it as a fluid with constant pressure, whereas the density $\rho$ relates to the barotropic index $\omega$ as $\rho \propto \omega^{-1}$ \cite{nocc}. Among the several ways to obtain this scenario, for example, it is possible to demand that the adiabatic index vanishes for a given barotropic fluid \cite{Dunsby16}, or alternatively to approach the scenario by means of a purely adiabatic fluid, whose sound speed identically vanishes. However, it has been argued that this model naturally leads to a zero sound speed and thus there is no need to
fix it by hand \emph{a priori} \cite{Luongo14,avi}. Although a dark fluid with constant pressure entirely matches the standard cosmological model, it is not able to describe inflation \cite{staro,Guth81,Linde82} and to fix the cosmological constant problem \cite{Weinberg89,Sahni02,Peebles03}. The cosmological constant problem states that it is not possible to remove, from the cosmological puzzle,  quantum vacuum energy whose fluctuations gravitate\footnote{The cosmological constant problem is based on two sub-issues: the discordance between predictions and observations of the cosmological constant, named fine-tuning problem, and the strange fact that matter and $\Lambda$ show very close magnitudes today, known as coincidence problem.}. 
A possibility would be to reduce the large value of $\Lambda$ through some mechanism of cancellation\footnote{All possibilities discussed in the literature take into account modifications or extensions of the whole description, complicating \emph{de facto} the standard cosmological paradigm.}. One may wonder whether the dark fluid can provide a mechanism of cancellation for $\Lambda$, overcoming the $\Lambda$CDM issues \cite{Luongo18}. For the sake of completeness, a key point of inflationary epoch is that the energy density is taken to be constant, \emph{i.e.} it cannot be reduced by space expansion. 

In this work, we propose a novel unified model that suggests how to unify dark energy and inflation by healing the cosmological constant problem. Our goal is to provide a picture in which the universe is accelerated by quantum fluctuations of vacuum energy and undergoes a phase transition from small to large scalar fields. During this phase, the universe is metastable and is fuelled by a double exponential potential (in analogy to solid state physics), acting to remove the cosmological constant term through a cancellation mechanism that extends the one proposed in \cite{Luongo18}. We interpret the emerging scalar field as a matter field with non-zero pressure. After the transition, the matter field behaves as a quasi-quintessence (QQ) scalar field, which significantly differs from quintessence because of the pressure definition\footnote{A similar approach has been developed in \cite{Gao10} and \cite{Lim10}.}. 
We study dynamics and thermodynamic properties of the QQ field, and then we analyze its role in removing the vacuum energy contribution.  
Under this scheme, we demonstrate that early and late times dynamics can be unified into a single scheme, if one invokes the QQ potential with  appropriate boundary conditions. These give rise to a Starobinsky-like potential, quite similar to the solid state physics Morse potential, with an inflationary phase characterized by attractors. The  stability analysis shows that our potential is chaotic, mimicking the  dark fluid after the transition. In this respect, our inflationary scenario  unifies the pictures provided by old and new inflation  into a single scheme. Here, the phase transition induces the mechanism responsible for removing vacuum energy, whereas chaotic inflation is responsible for the graceful exit from inflation. We finally discuss how to overcome the coincidence and fine-tuning problems and how to interpret our model as a dark fluid that mimics the $\Lambda$CDM model, fuelled however by a bare cosmological constant contribution. 

The paper is structured as follows. In \Cref{sezione2}, we provide the Lagrangian formulation of the QQ model. 
In \Cref{background}, the background cosmology of the new paradigm is shown, with a focus on its thermodynamics properties.
In \Cref{transition}, we explain the strategy to unify dark energy and inflation by means of a double exponential potential, where boundary conditions are imposed accordingly. 
The inflationary dynamics is then described in \Cref{sezione6}, in terms of stability, slow roll approximation and number of e-foldings. 
In \Cref{sezione7}, we discuss the theoretical consequences of our model. 
Finally, \Cref{sezione8} is dedicated to the final remarks and conclusions.

%%%%%%%%%%%%%%%%%%%%%%%%%%%%%%%%%%%%%%%%%%%%%%%%%%%%%%%%%%%%%%%%%%%%%%%%%%%%%%%%%%%%%%%%%%%%%%%%%%%%%%%%%%%%%%%%%%%%%%%%%%%%%%%%%%%%%%%%%%%%%%%%%%%%%%%%%%%%%%%%%%

\section{A quasi-Quintessence unified model}
\label{sezione2}

To construct the QQ model, we require the universe to accelerate at late times, guaranteeing structures to form at early times, and counterbalancing vacuum energy at primordial times. We thus seek for a unified dark energy model able to describe inflation and to heal the cosmological constant problem. Hence, we write down the main ingredients of the QQ scenario \cite{Luongo18}:
\begin{itemize}
\item[-]  a single fluid, represented by the QQ field, accounting for baryons and cold dark matter;
\item[-] dark matter with non-vanishing pressure, counterbalancing the quantum vacuum energy effects;
\item[-] pressureless baryons approximated by dust, as in the standard $\Lambda$CDM model;
\item[-] a \emph{first-order phase transition} is assumed to occur. The total energy is fixed through a Lagrange multiplier that, under suitable physical conditions, \emph{naturally} suggests an emergent constant negative pressure describing the universe after the transition;
\item[-] after the transition, the magnitude of the aforementioned pressure is comparable to the critical density. The corresponding scenario mimics the $\Lambda$CDM paradigm, but it represents \emph{de facto} a dark fluid model.  
\end{itemize}

\subsection{Lagrangian formulation}

The effective Lagrangian representation describing our QQ model is given by:
\begin{align}
\label{eq:1}
\mathcal{L} = K\left(X,\varphi\right) +\lambda\, Y\left[X,\nu\left(\varphi\right)\right]-V\left(X,\varphi\right)\,.
\end{align}
The kinetic functions $K$ and $Y$ and the Lagrangian multiplier $\lambda$, which  constraints the overall energy content of the universe, do not require any \emph{a priori} functional form.
The whole Lagrangian depends upon a scalar field $\varphi$ and its kinetic term $X \equiv g^{\alpha\beta}\nabla_\alpha \varphi \nabla_\beta\varphi/2$, being $g^{\alpha\beta}$ the metric tensor; the function $\nu(\varphi)$ plays the role of the specific inertial mass. 
The interacting potential $V$ is strictly requested to yield the first-order phase transition mentioned above\footnote{$V$ could be chosen arbitrarily, albeit the simplest one is a fourth-order symmetry-breaking potential.}.
From the variation of the action with respect to $\lambda$, $\varphi$ and $g^{\alpha\beta}$, we obtain, respectively,
\begin{align}
\label{eq:no5a}
&\,Y = 0\,, \\
\label{eq:no5b}
&\,\mathcal{L}_{,\varphi} - \nabla_\alpha \left( \mathcal{L}_{,X} \nabla^\alpha\varphi \right) = 0\,,\\
\label{eq:no6}
&\, T_{\alpha\beta}  = \mathcal{L}_{,X} \nabla_\alpha\varphi \nabla_\beta \varphi - \left(K -  V \right) g_{\alpha\beta}\,,
\end{align}
where the subscripts preceded by the comma label partial derivatives, so that $\mathcal{L}_{,X}=K_{,X}-V_{,X}+\lambda\, Y_{,X}$ and $\mathcal{L}_{,\varphi} = K_{,\varphi}-V_{,\varphi}+\lambda\, Y_{,\nu}\, \nu_{,\varphi}\,$.

Defining an effective $4$-velocity $v_\alpha \equiv \nabla_\alpha\varphi/\sqrt{2X}$, the energy-momentum tensor then reads
\begin{equation}
\label{eq:no10}
T_{\alpha\beta} = 2X \mathcal{L}_{,X} v_\alpha v_\beta - \left(K -  V \right) g_{\alpha\beta}\,,
\end{equation}
whose density and pressure become
\begin{align}
\label{eq:no11}
\rho =\, &2X \mathcal{L}_{,X} - \left(K -  V \right)\,,\\
\label{eq:no12}
P =\, &K - V\,.
\end{align}
As proved in \cite{Luongo18}, to guarantee structure formation, the pressure before and after the transition must be constant. As the potential $V$ takes constant (although different) values during these phases, it follows that $K=K_0={\rm const}$, being this constant the same before and after the transition, as we shall see in the next Section.
Bearing this in mind, we find
\begin{align}
\label{eq:no11bis}
\rho =\,&2X \mathcal{L}_{,X} + \mathcal V(\varphi)\,,\\
\label{eq:no12bis}
P =\, &- \mathcal V(\varphi)\,,
\end{align}
where  $\mathcal V(\varphi)\equiv V-K_0$. 
We note that $K$ taken as a constant does not imply neither $X$ nor $\mathcal L_{,X}$ to be constants. However, for the sake of simplicity, it is licit to fix $\mathcal L_{,X}$ to a given value, as we will discuss later on.

Within the framework of unified dark energy models, the QQ paradigm is not merely a slightly departing quintessence-like fluid, but rather it  can be considered
as an alternative version mainly different from  quintessence, written in terms of a ``new" modified energy momentum tensor. 
We shall explore better this point in what follows.

\subsection{General treatment of quasi-quintessence}

The general treatment, corresponding to the case in which $K$ is not a constant (during the transition phase), does not lead \emph{a priori} to a QQ model. 
The main differences in terms of density and pressure between the quintessence and the QQ approaches occur from the energy-momentum tensor shift
\begin{align}\label{tmunu}
 T_{\mu\nu}^{Q}-T_{\mu\nu}^{QQ}=\frac{\dot \psi^2}{2}\left(v_\mu v_\nu-g_{\mu\nu}\right)\,,
\end{align}
where conventionally we considered a generic scalar field, $\psi$, labelling both quintessence and QQ\footnote{The quintessence and the QQ scenarios are mostly different and cannot be described by the same field. The choice of taking the same field only lies on our desire to directly compare the two models.}. 
In principle, from the QQ model we can obtain quintessence by substituting $K\equiv \dot\psi^2/2$ and $2X\mathcal L_{,X}\equiv\dot\psi^2$, leading to 
\begin{subequations}
\begin{align}
 &\rho^{QQ}=\rho^{Q},   \label{rhoQ}\\
 &P^{QQ} =P^{Q}-\frac{\dot\psi^2}{2}\,, \label{pQ}
\end{align}
\end{subequations}
where we labelled with the superscripts QQ and Q our model and quintessence, respectively. 

Both quintessence and QQ provide a pressure that could be positive or negative. However, quintessence shows a pressure that can be positive or negative even if the potential sign is fixed, as a consequence of the fact that the pressure is given by the difference between the kinetic term and the potential itself. 
On the other hand, once the potential sign is fixed, QQ provides a pressure that does not depend upon the kinetic term and coincides with the potential itself, leading to the interesting feature of representing a unified description of dark energy and dark matter by means of a single field. 

It is worth to emphasize that our scenario is suitable to describe small perturbations and, in particular, structure formation, since it degenerates with the $\Lambda$CDM model and predicts a  vanishing speed of sound, differently from quintessence, for which $c_s^Q=1$.

The energy-momentum tensor suitable for our purposes provides the expression for pressure and density according to \Cref{rhoQ,pQ}. In order for \Cref{tmunu} to hold, we only consider the case $K= K_0={\rm const}$ after the transition. To reconcile the eras before and after the transition, one may postulate \cite{Gao10} 
\begin{equation}
T_{\mu\nu}\rightarrow 
T_{\mu\nu}+V (\psi) g_{\mu\nu}-\frac{1}{2}\nabla_\mu\psi\nabla_\nu\psi\,,
\end{equation}
for which the equations of motion for the scalar field are
\begin{equation}
\label{EoMSF}
\nabla^2\psi-2\frac{\partial V}{\partial \psi}+\dfrac{(\nabla^\mu\psi)(\nabla^\nu\psi)(\nabla_\mu\nabla_\nu\psi)}{\nabla_\alpha\psi\nabla^\alpha\psi}=0\,,
\end{equation}

\noindent where we restored the information $\psi=\psi(t,{\bf x})$, instead of considering the time-dependence only.

\section{Background cosmology}
\label{background}

Our idea is to unify the dark sector, so that the effects of the dark energy can be unveiled by
means of the dark matter, leaving baryons as the unique fluid behaving as pure pressureless matter.
This approach slightly differs from the one developed in \cite{Luongo18}, where the dark energy was proposed as intimately connected to the pressure term of baryonic matter.
Here, dark energy, modelled by a scalar field obeying \Cref{EoMSF}, is the result of the difference between the negative pressure of dark matter (coinciding with $K_0$ before the transition) and the offset of the effective potential, related to the vacuum energy density. Therefore, the dark fluid enters the energy-momentum tensor as a source for the net energy content of the universe and provides a negative pressure that dominates over matter at late times \cite{Linder09}. 

To show that, we focus on a flat Friedmann-Robertson-Walker (FRW) metric 
$ds^2=-dt^2+a^2(t)\delta_{ij}dx^idx^j$, where $a(t)$ is the scale factor, which is nowadays $a_0=1$. 
Thus, the density and the pressure of the field $\varphi$ become
\begin{align}
\label{endensphi}
&\rho_\varphi=\mathcal L_{,X}\dot{\varphi}^2+\mathcal V(\varphi)\,,\\
&P_\varphi=-\mathcal V(\varphi)\,.
\end{align}
where the ``dot'' denotes the  derivative with respect to the cosmic time.
In the Friedmann equations, the cosmic fluid can be split into matter and scalar field densities
\begin{subequations}
\begin{align}
\label{Fr1}
&3H^2=\kappa^2\left(\rho_b+\rho_\varphi\right),\\
\label{Fr2}
&2\dot{H}+3H^2=-\kappa^2 P_\varphi\,,
\end{align}
\end{subequations}
and one can define the normalised densities
\begin{equation}
\Omega_b\equiv\dfrac{\kappa^2\rho_b}{3H^2}\ , \quad  \Omega_\varphi\equiv\dfrac{\kappa^2\rho_\varphi}{3H^2}= 1-\Omega_b \,,
\end{equation}
where $\kappa^2\equiv 8\pi/M_{\rm Pl}^2$ with $M_{\rm Pl}$ being the Planck mass.
From \Cref{Fr1,Fr2}, we obtain the continuity equation for baryons and the equation of motion of the QQ field, respectively,
\begin{align}
&\dot{\rho}_b+3H\rho_b=0\,,\\
\label{phieqofcont}
&\ddot{\varphi}+\dfrac{3}{2}H\dot{\varphi}+\frac{\mathcal V_{,\varphi}(\varphi)}{2\mathcal L_{,X}}=0\,.
\end{align}

\subsection{Thermodynamic interpretation of the quasi-quintessence field}

Barotropic models differ from quintessence for the corresponding sound speed. For instance, from the one hand a given quintessence potential leads to a pressure that is not uniquely defined in terms of the density. From the other hand, barotropic fluids provide pressures that are function of the density only.
Therefore, even if one can postulate a quintessence model depending upon single-valued pressure being function of the density, to distinguish it from barotropic fluid one has to compute the sound speed. 

In the case of quintessence, the sound speed is $c_{s}^{Q}=1$, whereas for barotropic fluids we have $c_s\leq 1$. 
To ensure stability, the sound speed might be positive-definite. Also, barotropic models and scalar field models occupy disjoint regions in the $\omega-\omega^\prime$ space, where the ``prime'' denotes the derivative with respect to the number of \textit{e-foldings} $N\equiv \ln a$. 
The general solution for a constant sound speed is then given as
\begin{equation}\label{equaz4w}
\omega^\prime+3(c_s^2-\omega)(1+\omega)=0\,,
\end{equation}
where the case $c_s^2=1$ represents the only exception when scalar field models and barotropic dark energy provide the same predictions.

As a consequence of the above considerations, for a constant sound speed in the Zeldovich interval, we have
\begin{equation}
\rho=A+B\,a^{-3(1+c_s)}\,,
\end{equation}
which reduces to $\rho\propto a^{-6}$ for quintessence models, \emph{i.e.} when the scalar field substitutes a barotropic fluid\footnote{This case has been investigated also in the framework of the K-essence models \cite{Scherrer04,Armendariz99,Chiba00} and is similar to several other unified scenarios, among which the Chaplygin gas model.}.
On the other hand, a fully-degenerate case occurs for \Cref{rhoQ,pQ}, where one has
\begin{equation}\label{soundo}
c_s^{QQ}\neq c_s^{Q}\,,  \quad  c_s^{QQ}=0\,, 
\end{equation}
appearing indistinguishable from the $\Lambda$CDM model, although matter arises as an additional parameter. 

The QQ model is also different from K-essence as $P\neq f(X)$, so that the two scenarios do not degenerate between them\footnote{This may not be the case during the transition, because $K$ may not be a constant and, thus, $P=f(X)$.}. The trajectories for kinetic K-essence span over the $\omega - \omega^\prime$ space, analogously to those of barotropic fluids.

Hence, the QQ paradigm provides a strategy to heal the cosmological constant problem, assuming the existence of a single fluid composed by matter only, under the form of baryons
and cold dark matter. However, in this scenario, cold dark matter exhibits a non-vanishing pressure, which decelerates the vacuum energy-dominated universe and cancels out the contributions of quantum fluctuations. 
Concerning the last point, we note that for dark matter and/or baryons considered as collisional particles, the net pressure can be different from zero. The dust case, in fact, occurs when the mass of the single constituent for matter, namely a proton, a dark matter particle, etc., is larger than the environment temperature. Indeed, the equation of state for a perfect gas reads $P=nk_BT/m_X$, with $n$ being the total number of particles, $k_B$ the Boltzmann constant, while $T$ and $m_X$ are the absolute temperature and the constituent mass, respectively. Dust means $m_X\gg T$, and this occurs as the temperature decreases, but not at all stages of the universe's evolution. 

In the present study, we consider the QQ field $\varphi$ as an effective representation of dark matter with pressure, whose total  energy is limited by means of  Lagrange multipliers. Consequently, we assume dark matter and baryons to evolve according to distinct equations throughout the cosmic evolution, distinguishing their dynamics from a relativistic point of view. 
Then, including a phase transition induced by a particular form of the symmetry-breaking potential, we can find that the corresponding (dark) matter dynamics is described by a perfect, irrotational and isentropic fluid, characterized by a positive Helmotz energy, which \emph{naturally} exhibits a negative pressure \cite{Luongo18}.
Therefore, the non-zero pressure term of the QQ model provides the late-time cosmic dynamics due to the presence of a bare cosmological constant.

%%%%%%%%%%%%%%%%%%%%%%%%%%%%%%%%%%%%%%%%%%%%%%%%%%%%%%%%%%%%%%%%%%%%%%%%%%%%%%%%%%%%%%%%%%%%%%%%%%%%%%%%%%%%%%%%%%%%%%%%%%%%%%%%%%%%%%%%%%%%%%%%%%%%%%%%%%%%%%%%%%

\section{Unifying inflation with phase transition}
\label{transition}

As claimed above, we require a symmetry-breaking phase-transition potential in order to cancel out the vacuum energy quantum fluctuations. A prototype of such a potential is a fourth-order potential, albeit this is not the unique possibility. Several other phase-transition potentials can be used, providing different behaviors \cite{Guth81,Linde_super}. For convenience, we split our description in two different phases: before/after and  during the transition.

\subsection{Before and after the transition}
\label{sezione3}

We shall now address the cosmological constant problem, invoking the form of $V$. In \cite{Luongo18} we described the phase transition through a symmetry-breaking (thermalized) fourth-order potential of the form
\begin{equation}
\label{eq:pottemperature}
V(\varphi)=V_0+\frac{\chi}{4}\left(\varphi^2-\varphi_0^2\right)^2+\frac{\chi}{2}\varphi_0^2\varphi^2
\left(\frac{T}{T_{\rm c}}\right)^2\,,
\end{equation}
where $\chi$ is a dimensionless coupling constant, $\varphi^2_0$ is the value of $\varphi$ at the minimum of the potential, $T$ is the temperature, and $T_{\rm c}$ is the critical temperature setting the beginning of the transition. We argue that:

\begin{itemize}
    \item[-] before the transition, when $T>T_{\rm c}$, the minimum of $V$ is located at $\varphi=0$, \emph{i.e.} $V_0+\chi \varphi_0^4/4$;
    \item[-] during the transition, inflation occurs, and the vacuum energy contributes as source for a de Sitter accelerating phase;
    \item[-] after the transition,  when $T<T_{\rm c}$, the minimum is $V_0$ for $\varphi=\varphi_0$.
\end{itemize}

The cosmological constant problem is intimately related to the value of the potential offset $V_0$. The issue arises whether we require the  vacuum energy to vanish before
or after the transition. 
If the vacuum energy density $\rho_{vac}$ is set to zero before transition by choosing $V_0 = -\chi\varphi_0^4/4$, then $\rho_{vac}\neq0$ must hold afterwards. 
Conversely, if vacuum energy is set to zero after the transition by choosing $V_0 = 0$, thus before the transition we have $\rho_{vac}\neq0$. 
Anyhow, we notice that vacuum energy cannot be zero before and after the transition as well, since the offset $V_0$ cannot vanish in those periods.

Within the QQ model, choosing $V_0 = -\chi\varphi_0^4/4$, we cancel the vacuum energy by simply having:
\begin{itemize}
\item[-] before the transition: $\mathcal V=-K_0$, $\rho=2X\mathcal L_{,X}-K_0$ and     $P=K_0$;
\item[-] after the transition:    $\mathcal V = -K_0 -\chi\varphi_0^4/4$, $\rho=2X\mathcal L_{,X}-K_0-\chi\varphi_0^4/4$ and $P=K_0+\chi\varphi_0^4/4$.
\end{itemize}
The cancellation occurs as we have $-K_0-\chi\varphi_0^4/4\simeq \rho_{cr}$, where $\rho_{cr}\equiv 3H_0^2/\kappa^2$ is the universe's critical density. 
Summarizing: 

\begin{itemize}
    \item[-] the QQ model is capable of describing the universe evolution immediately after the Big Bang;
    \item[-] assuming a phase transition, the model is able to heal the cosmological constant problem if the potential is constant before and after the transition;
    \item[-] the function $K$ is kept constant before and after the transition, as a consequence of the thermodynamics of matter with non-zero pressure \cite{Luongo18}. 
\end{itemize}

It is important to stress that \Cref{eq:pottemperature} is valid only before and after the transition. During the transition, the ponential should have a more complicated form in order to address the complexity of the corresponding metastable phase. 
Moreover, the fourth-order potential has been recently ruled out from the candidate inflationary potentials \cite{Planck}. Since we attempt to describe inflation during the transition, we can physically construct a potential that differs from \Cref{eq:pottemperature}, but showing  a symmetry-breaking behavior that provides a phase transition with the properties described above.

%%%%%%%%%%%%%%%%%%%%%%%%%%%%%%%%%%%%%%%%%%%%%%%%%%%%%%%%%%%%%%%%%%%%%%%%%%%%%%%%%%%%%%%%%%%%%%%%%%%%%%%%%%%%%%%%%%%%%%%%%%%%%%%%%%%%%%%%%%%%%%%%%%%%%%%%%%%%%%%%%%

\subsection{During the transition}

The requirements to build a potential in the metastable phase, when the universe undergoes the transition, are: 
\begin{itemize}
    \item[-]  the potential should induce a phase transition, \emph{i.e.} its limit for small fields is $\propto\varphi^4$; 
    \item[-] we need to escape the transition, \emph{i.e.} once the metastable phase ends, the universe exits the transition lying on the symmetry minimum. 
\end{itemize}

Physically speaking, the above requirements resemble the inflationary behavior. Thus, we assume inflation to be driven by vacuum energy during the metastable phase. Consequently, the inflaton field is fuelled by the vacuum energy density and the graceful exit could arise once the vacuum energy is removed by means of the aforementioned mechanism.
Thus, some considerations follow.
\begin{itemize}
\item [-] The phase transition is induced by the field with a potential that is slightly different from the fourth-order one.
\item[-] The transition is clearly induced by the field itself. This plays the role of the inflaton during the metastable phase, namely during the transition.
\item[-] Vacuum energy is suppressed during the transition through the cancellation mechanism proposed by means of the QQ field.
\item[-] The above mechanism appears similar to \emph{old inflation}, being characterized by a phase transition from the very beginning. However, the potential might escape the metastable phase alternatively to the graceful exit scenario, and thus the potential could be formulated to reach a fixed point as $\varphi\rightarrow\varphi_0$ (the minimum) or better $\dot \varphi\rightarrow0$ as $\varphi\rightarrow\varphi_0$, meaning that all the kinetic energy of the field has been employed in particle creation due to damped oscillations around the minimum.
\item[-] The latter  prerogative is essentially analogous to the \emph{new inflation}. Hence, we expect to unify the main features of old and new inflation into a single scheme. In other words, we propose a new inflation induced by the old inflation mechanism, namely the transition.
\end{itemize}

Since during inflation the potential might reach a fixed point, we assume that the phase transition is due to vacuum energy and stops once an attractor is reached with conditions  $\varphi\rightarrow\varphi_0$ and $\dot \varphi\rightarrow0$.
After the transition, the vacuum energy contribution changes accordingly to the aforesaid scheme, \emph{i.e.} it likely transforms to particles created during reheating phase \cite{Dolgov} and/or by particle production due to geometry \cite{GravBar}. 

Hence, a suitable potential can be constructed as: we assume the potential to be smoothly connected to the phases before and after transition to guaranteeing continuity in the energy budget of the universe; the potential should be compatible with current observations \cite{Planck}, and the coupling constant $\chi$ should resemble the one induced the vacuum energy; the thermodynamics of the potential should exhibit a phase transition of the first order.

\subsection{Inflationary potential}
\label{sec:potentials}

In view of the above requirements, we consider a positive-definite potential over the whole phase space. Thus, we start by writing the most general double-exponential form:
\begin{equation}
    \label{pot_prelimin}
    V(\Phi) = V_0+\mathcal A\left[a\exp(\alpha\Phi^{c_1})+b\exp(\beta\Phi^{c_2})\right]^m\,,
\end{equation}
where  $\Phi\equiv\varphi/\varphi_0$ is the QQ field normalized over its value $\varphi_0$ at the minimum of the potential\footnote{Such normalization does not introduce a divergence since in our model, after the transition, the $\varphi$ field evolves towards the minimum of the potential, which always occurs at $\varphi_0\neq0$.}.
This choice is motivated by the fact that the arbitrary constants $\mathcal A$, $a$, $b$, $\alpha$, $\beta$, $c_1$, $c_2$ and $m$ can be fixed in a general and unrestricted manner as follows.
Requiring $V(0)=V_0+\chi\varphi_0^4/4$ at $\Phi=0$ (or $\varphi=0$), implies that 
\begin{equation}
    a=-b \exp(\beta - \alpha)\,.
\end{equation}
Also, for $\Phi=1$ (or $\varphi=\varphi_0$) we must have $V(1)=V_0$, leading to
\begin{equation}
\mathcal A = 
    (\chi\varphi_0^4/4) \left[b - b\exp(\beta -\alpha)\right]^{-m}\,.
\end{equation}
Moreover, to enable spontaneous baryogenesis, in the regime of small oscillations around $\Phi=1$, the potential has to be quadratic \cite{Dolgov}, which implies that $m=2$. 

\begin{figure*}
    \centering
    a)
    \includegraphics[width=0.45\linewidth]{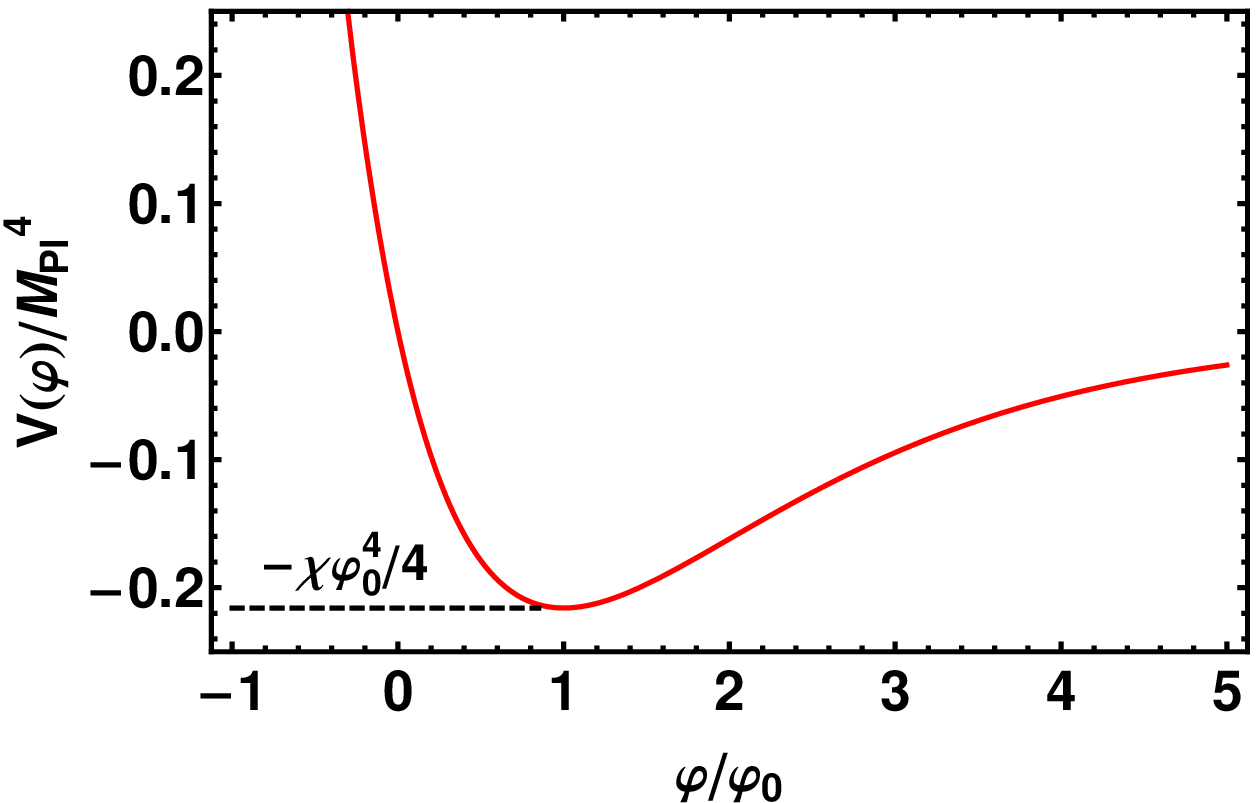}
    \hfill
    b)
    \includegraphics[width=0.45\linewidth]{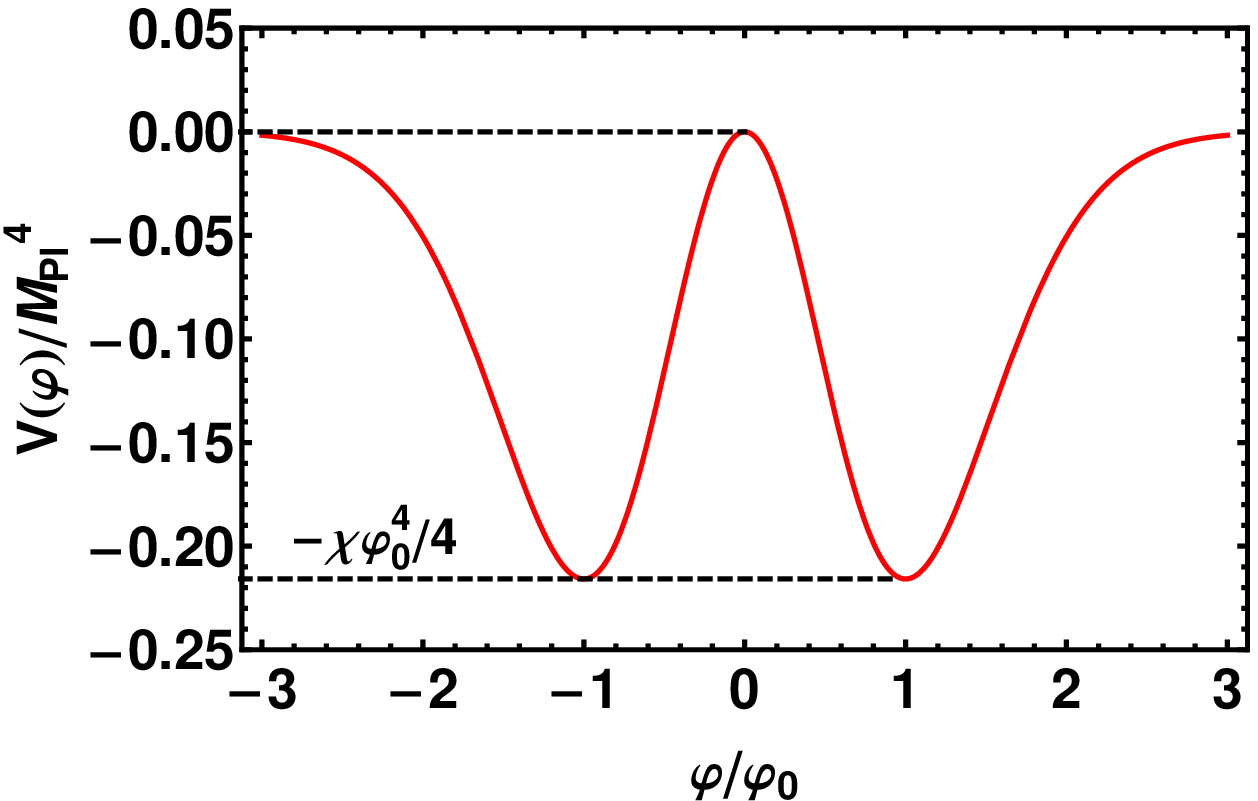}\\
    \vspace{0.5cm}
    c)\includegraphics[width=0.45\linewidth]{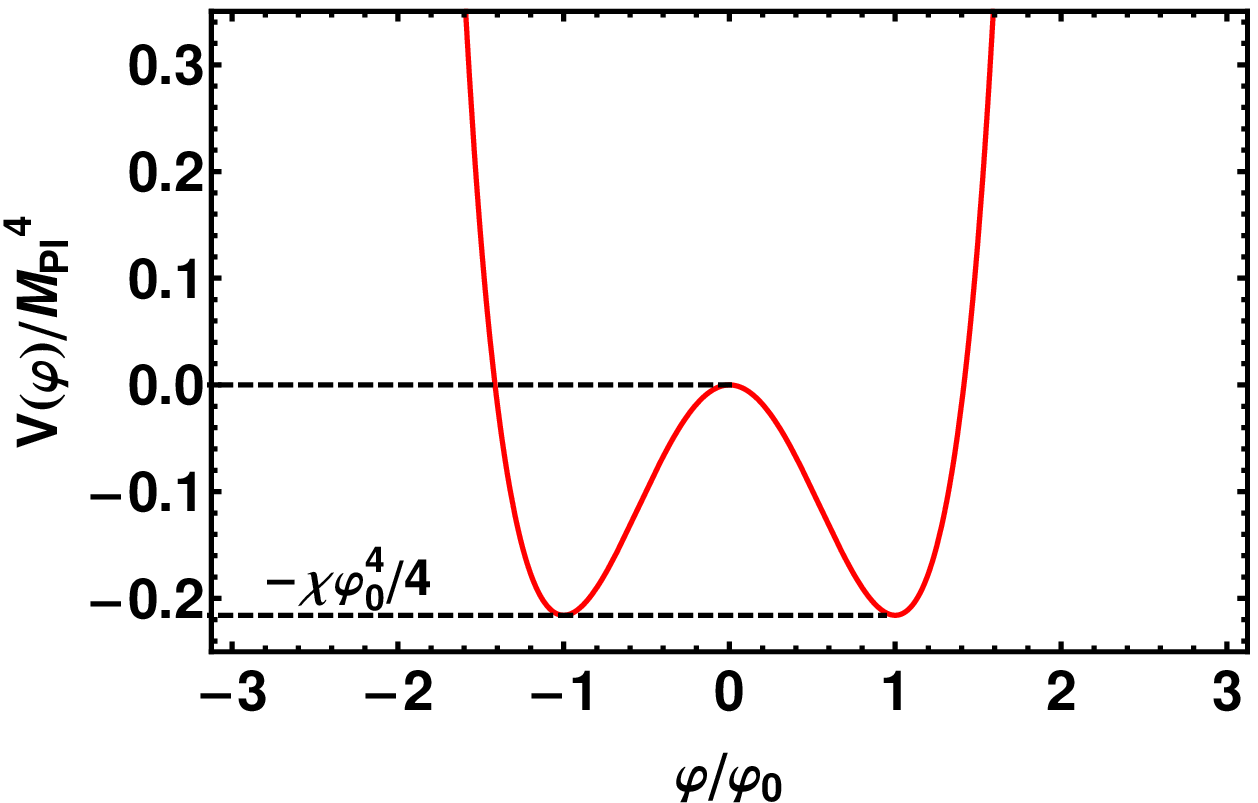}
    \hfill
    d)
    \includegraphics[width=0.45\linewidth]{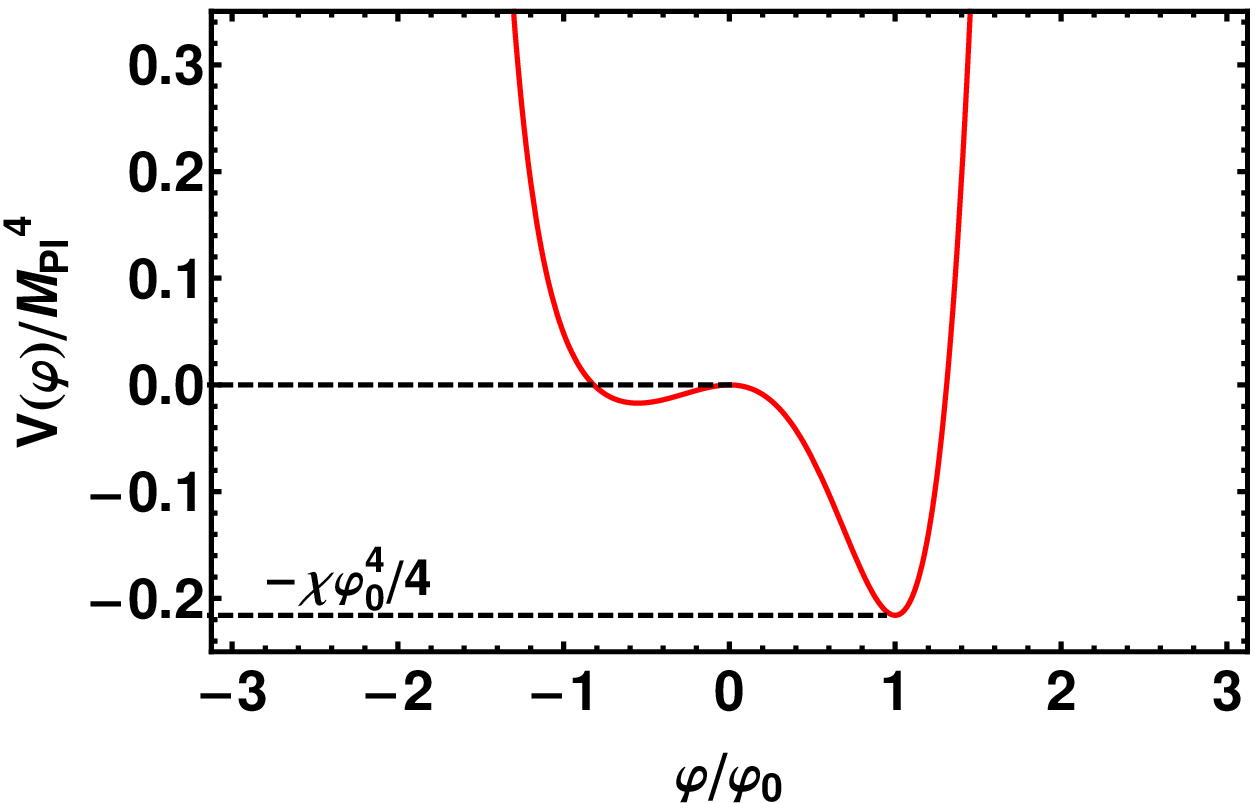}
    \caption{Inflationary potential discussed in \Cref{sec:potentials}: a) the Starobinsky-like in \Cref{stalike}; b) the symmetry-breaking-like with finite potential walls in \Cref{symbreakwrong}; c) the symmetry-breaking-like with infinite potential walls in \Cref{symbreaktrue}; d) the false vacuum case extracted from \Cref{intermediateV}. The values of the constants employed in the plots are discussed in the text.}
    \label{Potentials}
\end{figure*}

For $\beta=0$, $\alpha<0$ and $c_1=1$, we thus obtain
\begin{equation}
\label{stalike}
V_1(\Phi) = V_0+\frac{\chi\varphi_0^4}{4} \left[\frac{1-e^{-|\alpha|(\Phi-1)}}{1-e^{|\alpha|}}\right]^2\,,
\end{equation}
mimicking the features of the Starobinsky potential $V(\Phi)=\Lambda^4[1-e^{-\sqrt{2/3}(\varphi_0/M_{\rm Pl})\Phi}]$ \cite{staro} for
\begin{equation}
\Lambda^4\simeq \frac{\chi\varphi_0^4}{4}\left(1-e^{|\alpha|}\right)^{-2}\,,\quad \alpha\simeq \sqrt{\frac{2}{3}}\frac{\varphi_0}{M_{\rm Pl}}\,.
\label{alphasta}
\end{equation}
Requiring $V_1(0)= V_1(+\infty)$ and $V_0=-\chi\varphi_0^4/4$, we find the constraint $|\alpha|=\ln 2$. Further, plugging in this value in the expression of $\alpha$ given by \Cref{alphasta}, we obtain $\varphi_0/{\rm M_{Pl}}= \sqrt{3/2}\ln 2\approx0.85$.
The behavior of the Starobinsky-like potential \eqref{stalike} is shown in \Cref{Potentials}a, where we set the vacuum energy density $\epsilon_{\rm v}=-V_0=\chi\varphi_0^4/4=0.22\,M_{\rm Pl}^4$, so that $\chi=1.66$.

An interesting case corresponds to $\beta=0$, $\alpha<0$ and $c_1=2k$, with $k\geq1$. Here, the potential has two symmetric minima and it provides a symmetry breaking. In fact, without loss of generality, setting $k=1$ we find
\begin{equation}
\label{symbreakwrong}
V_2(\Phi) = V_0+\frac{\chi\varphi_0^4}{4} \left[\frac{1-e^{-|\alpha|(\Phi^2-1)}}{1-e^{|\alpha|}}\right]^2.
\end{equation}
In this model, unlike the fourth-order potential, the energy necessary to escape from the minimum is finite.
It is worth noting that also for the $T$-model potential \cite{Tmodel} both walls are finite.
The inflaton is treated as a classical field and, around the minimum $\varphi_0$, it experiences the Hubble friction term that dissipates the initial energy and dampens the oscillations.
Within this qualitative picture, the inflaton does not posses the necessary energy to escape from the minimum, and one may think to fix the height of the potential walls imposing $V_2(0)= V_2(\pm\infty)$. This greatly simplifies the complexity of the potential, leading again to $|\alpha|= \ln 2$.
This $W$-shaped potential, dubbed as $W$-model, is shown in \Cref{Potentials}b, where we set, in analogy to the previous case, $\epsilon_{\rm v}=-V_0=\chi\varphi_0^4/4=0.22\,M_{\rm Pl}^4$.

Instead, by considering $\beta\neq0$, we obtain
\begin{equation}
\label{intermediateV}
    V_3(\Phi) = V_0+\frac{\chi\varphi_0^4}{4} \left[\frac{e^{\beta\Phi^{c2}}-e^\beta e^{\alpha(\Phi^{c1}-1)}}{1-e^{-\alpha+\beta}}\right]^2,
\end{equation}
and, for $\alpha<0$, $|\alpha|=\beta$ and $c_1=c_2=2$, we have
\begin{equation}
\label{symbreaktrue}
    V_3(\Phi) = V_0+\frac{\chi\varphi_0^4}{4} e^{2|\alpha|\Phi^{2}}\left[\frac{1-e^{-2|\alpha|(\Phi^{2}-1)}}{1-e^{2|\alpha|}}\right]^2
    \,.
\end{equation}
The corresponding shape is displayed in \Cref{Potentials}c, where we set again $\epsilon_{\rm v}=-V_0=\chi\varphi_0^4/4=0.22\,M_{\rm Pl}^4$ and choose $|\alpha|=\beta=0.5$.

It is worth to mention that in \Cref{intermediateV} the requirements $|\alpha|=\beta$ and $c_1=c_2$ are crucial: arbitrary values may lead to nonphysical potentials and/or complicated behaviors, including the existence of false vacua. A broken symmetry system in a false vacuum state implies that the vacuum of the universe is tunnelling through the real vacuum state. 
To keep our treatment simple, we do not consider the thorny cases $\beta\neq|\alpha|$ and $c_1\neq c_2$, avoiding thus any false vacua.
This situation is shown in \Cref{Potentials}d where we set once again $\epsilon_{\rm v}=-V_0=\chi\varphi_0^4/4=0.22\,M_{\rm Pl}^4$ and choose $\alpha=-0.3$, $\beta=0.5$, $c_1=3$ and $c_2=2$.

An alternative functional form for the above potentials, which makes more explicit their dependence upon the field $\varphi$ but $\alpha$ and $\beta $ dimensional, is easily attainable by setting $\Phi\rightarrow\varphi/\varphi_0$, $\alpha\rightarrow\alpha\varphi_0^{c_1}$ and $\beta\rightarrow\beta\varphi_0^{c_2}$.
This representation has the advantage of providing constraints on the value of $\varphi_0$ from the analysis of the inflationary dynamics. 
In the specific case of the $W$-model, which will be employed in the following sections,
this can be obtained by setting $\Phi\rightarrow\varphi/\varphi_0$ and $|\alpha|\rightarrow|\alpha|\varphi_0^2$, leading to
\begin{equation}
\label{symbreakwrong2}
V_2(\varphi) = V_0+\frac{\chi\varphi_0^4}{4} \left[\frac{1-e^{-|\alpha|(\varphi^2-\varphi_0^2)}}{1-e^{|\alpha|\varphi_0^2}}\right]^2.
\end{equation}

%%%%%%%%%%%%%%%%%%%%%%%%%%%%%%%%%%%%%%%%%%%%%%%%%%%%%%%%%%%%%%%%%%%%%%%%%%%%%%%%%%%%%%%%%%%%%%%%%%%%%%%%%%%%%%%%%%%%%%%%%%%%%%%%%%%%%%%%%%%%%%%%%%%%%%%%%%%%%%%%%%

\section{Inflationary dynamics}
\label{sezione6}

In this section, we study the inflationary dynamics of the QQ model in the slow-roll approximation. Specifically, substituting \Cref{endensphi} in \Cref{Fr1}, we find 
\begin{equation}
\label{1FriedSlow}
3H^2 = \kappa^2
\left[\mathcal L_{,X} \dot{\varphi}^2 + \mathcal V(\varphi) \right].
\end{equation}
It is worth mentioning that the above relation is general and not limited to the inflationary epoch. Indeed, we did not assume \emph{a priori} any approximations on \Cref{phieqofcont} and the above first Friedmann equation can be adapted to any epoch, before and after inflation, as one can see in Fig. \ref{fig:chaotic}. In this section, we limit to the inflation era so that we can impose the slow roll conditions over the field and the potential, namely
$\mathcal L_{,X}\dot{\varphi}^2 \ll \mathcal V(\varphi)$ and $\ddot{\varphi} \ll 3H\dot{\varphi}/2$.
Hence, \Cref{phieqofcont,1FriedSlow} become
\begin{align}
3H^2 &\simeq \kappa^2 \mathcal V(\varphi)\,,
\label{2_3_4}\\
    3H\dot{\varphi}&\simeq - \frac{\mathcal V_{,\varphi}(\varphi)}{\mathcal L_{,X}}\,.
\label{2_3_5}
\end{align}
We thus introduce the slow-roll parameters
\begin{align}
\epsilon &\equiv -\frac{\dot H}{H^2}\simeq
\frac{1}{2\kappa^2\mathcal L_{,X}}\left(\frac{\mathcal V_{,\varphi}}{\mathcal V}\right)^2,\\ 
\eta &\equiv -\frac{\ddot \varphi}{H\dot\varphi} -\frac{\dot H}{H^2} \simeq
\frac{1}{\kappa^2\mathcal L_{,X}} \left(\frac{\mathcal V_{,\varphi\varphi}}{\mathcal V}\right),
\label{2_3_6}
\end{align}
satisfying the conditions $\epsilon,|\eta|\ll1$ at the beginning of inflation, whereas $\epsilon=|\eta|=1$ at the end of inflation.

We choose to focus our analysis on the $W$-model potential in the form given by \Cref{symbreakwrong2}, as such a potential seems to reconcile the spontaneous symmetry breaking and the chaotic inflation pictures. 
Moreover, the $W$-model represents a minimal extension of the Starobinsky potential, which is one of the best candidates for inflation as of today \cite{Planck}. For simplicity, we fix $\mathcal L_{,X}=1$ and assume that inflation occurs at $\varphi\ll1$.
Under this assumption, we obtain\footnote{We initially expand around $\varphi=0$ up to the first order only to determine the scale factor evolution. In the following, we consistently expand up to the second order.} %
\begin{align}
\label{apprepsW}
\epsilon &\approx \frac{32 (\ln2)^2}{\kappa^2}\frac{\varphi^2}{\varphi_0^4}\,,\\ 
\label{appretaW}
\eta &\approx -\frac{8\ln2}{\kappa^2\varphi_0^2} + \frac{40 (\ln2)^2}{\kappa^2}\frac{\varphi^2}{\varphi_0^4}\,.
\end{align}

As we require the potential to behave, at the same time, as a spontaneous-symmetry-breaking and chaotic inflation, we assume as a necessary condition (like in the chaotic inflation scenario) that $\varphi_{\rm end}$ is the same as computed from the conditions $\epsilon=|\eta|=1$. 
Thus, from \Cref{apprepsW,appretaW}, we obtain the constraint on $\varphi_0$:
\begin{equation}
\label{apprphi0}
\varphi_0 \simeq 4\sqrt{\frac{2\ln2}{\kappa^2}} \approx 0.94 M_{\rm Pl}\,.
\end{equation}
To work out the initial and final settings of inflation, we evaluate the approximate solution for $\varphi(t)$ and $a(t)$ from \Cref{1FriedSlow,2_3_4,2_3_5}, bearing in mind that $H\equiv\dot a/a$:
\begin{align}
\varphi(t)&\simeq \varphi_i \exp\left(4\ln 2\sqrt{\frac{\chi}{3\kappa^2}}t\right)\,,\\
a(t)&\simeq a_i \exp\left(16\ln 2\sqrt{\frac{\chi}{3\kappa^2}}t\right),
\end{align}
where $\varphi_i$ and $a_i$ are the values of the field and the scale factor, respectively, at the beginning of inflation. 

Setting $\epsilon=1$ in \Cref{apprepsW} and using \Cref{apprphi0}, we find
\begin{equation}
\label{apprphiend}
\varphi_{\rm end} \simeq \frac{\varphi_0}{\sqrt{\ln 2}} \approx 1.13 M_{\rm Pl}\,.
\end{equation}
The values of $\varphi_0$ and $\varphi_{\rm end}$ computed in \Cref{apprphi0,apprphiend}, respectively, break the approximation $\varphi\ll1$ invoked above.
However, this is not surprising, because both $\varphi_0$ and $\varphi_{\rm end}$ have been computed from the requirements $\epsilon=1$, \emph{i.e.} when the slow-roll $\epsilon\ll1$ condition is broken.
Moreover, in view of the above approximations, \Cref{apprphi0,apprphiend} imply that $\varphi_{\rm end}>\varphi_0$. 
Since inflation cannot finish at values $\varphi>\varphi_0$, we use a further condition on the end of the inflation, namely $\varphi_f\equiv \min[\varphi_{\rm end},\varphi_0]$. 

In this regard, the flatness problem occurs as the final total density, $\Omega_f$, is limited as  $|\Omega_f-1|\sim10^{-60}$ right after the end of inflation. Baptizing the initial total density of the universe by $\Omega_i$, the ratio between the initial and the final phase of inflation reads
\begin{equation}\label{ratiodensities}
\frac{|\Omega_f-1|}{|\Omega_i-1|} \simeq \left(\frac{a_i}{a_f}
\right)^2=e^{-2N}\,.
\end{equation}
It is worth noticing that  $\Omega_{i;f}-1\equiv\Omega_k$, with $\Omega_k$ being the spatial curvature of the universe. Thus, assuming $|\Omega_i-1|\simeq1$, to solve the flatness problem we require $N\simeq70$ and, in analogy, a quite similar number of e-foldings is required to solve the horizon problem. We note that the symbol $\simeq$ in \Cref{ratiodensities} is due to the approximation $|\Omega_i-1|\simeq1$, whereas the functional dependence on the right side, i.e., $\propto a^2$, is due to how $\Omega_k$ scales as a function of the scale factor.

Taking into account our potential, in the slow-roll approximation we have 
\begin{equation}
N=\sqrt{\frac{\kappa^2}{8}}\int_{\varphi_i}^{\varphi_f}\frac{d\varphi}{\sqrt{\epsilon}}\simeq\frac{\kappa^2\varphi_0^2}{16\ln 2}\ln\left(\frac{\varphi_f}{\varphi_i}\right),
\end{equation}
which requires $\varphi_i\approx 5.92\times10^{-16}$ M$_{\rm Pl}$. This value is very close to zero, meaning that inflation starts at $\varphi\ll1$, as assumed above.

\section{Stability of the model}

Now, we investigate the stability of the QQ model. 
The aim of this analysis is to prove the existence of stable points, where the inflation stops once the conditions $\epsilon=|\eta|=1$ are attained, which represents an attractor solution of the model as well.
We work out dimensionless variables defined as in \cite{D'Agostino18} 
\begin{equation}
x\equiv\dfrac{\kappa\sqrt{\mathcal L_{,X}}\dot\varphi}{\sqrt{3}H}\,,\quad y\equiv\dfrac{\kappa\sqrt{\mathcal V}}{\sqrt{3}H}\,,
\end{equation}
which defines the constraint
\begin{equation}
x^2+y^2+\Omega_b=1\,.
\end{equation}
Here, $\Omega_b$ is the baryon density parameter and we have
\begin{subequations}
\begin{align}
&\Omega_\varphi=x^2+y^2\ ,\\
&w_\varphi\equiv\dfrac{p_\varphi}{\rho_\varphi}=-\dfrac{y^2}{x^2+y^2}\ .
\end{align}
\end{subequations}

The dynamical equations can be thus written as
\begin{subequations}
\begin{align}
x'&=-\dfrac{3}{2}xy^2+\sqrt{\dfrac{3}{2}}\,\epsilon\, y^2\,,\\
y'&=\dfrac{3}{2}y(1-y^2)-\sqrt{\dfrac{3}{2}}\, \epsilon\, xy\,.
\end{align}
\end{subequations}
Defining the set $\mathbf{X}=(x,y)$, where $x\geq0$ and $y\geq0$, it is easy to compute the critical points satisfying the differential equation $\mathbf{X}'=0$:
\begin{align}
&\mathbf{X}_c^{(1)}=(x\,, \, 0)\,, \label{first critical point background}\\
&\mathbf{X}_c^{(2)}=\left(\sqrt{\dfrac{2}{3}}\,\epsilon\,,\ \sqrt{1-\dfrac{2\epsilon^2}{3}}\right) . \label{second critical point background}
\end{align}
In correspondence of the first critical point, one has
\begin{equation}
\Omega_\varphi^{(1)}=x^2 \quad,\quad w_\varphi^{(1)}=0\,.
\end{equation}
Since inflation ends when $\epsilon=1$, the second critical point exists for $0<\epsilon\leq1$ and one finds
\begin{equation}
\Omega_\varphi^{(2)}=1 \quad,\quad w_\varphi^{(2)}=-1+\dfrac{2\epsilon^2}{3}\,.
\end{equation}

\begin{figure}
\begin{center}
\includegraphics[width=0.98\linewidth]{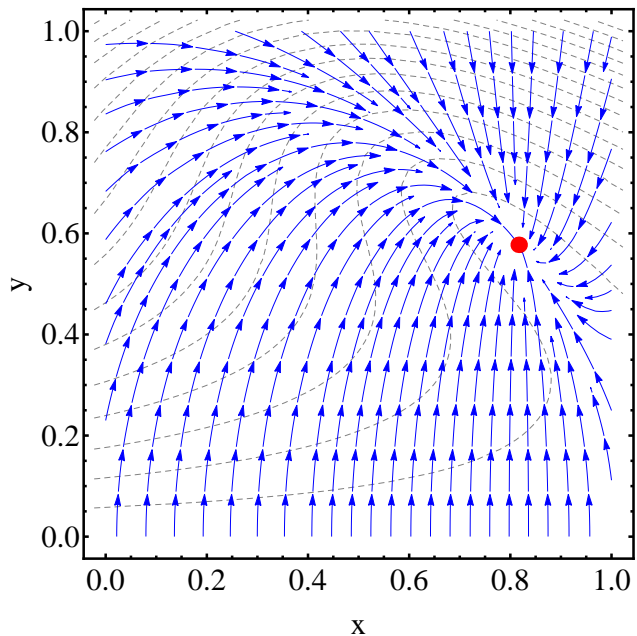}
\caption{Phase-space diagram at the background level for $\epsilon=1$. The red dot denotes the attractor point of the dynamical system (cf. \Cref{second critical point background}), the blue arrows depict the vector field, and the dashed gray lines indicate level curves.}
\label{fig:background attractor}
\end{center}
\end{figure}

The stability of the critical points is accounted by demanding $\mathbf{X}=\mathbf{X}_c+\mathbf{\delta X}$, where $\mathbf{\delta X}=(\delta x, \delta y)$. Thus, linear perturbations of the dynamical variables are obtained from $\delta \mathbf{X}'=\mathcal{M}\ \delta\mathbf{X}$, where
\begin{equation}
\mathcal{M}=
\begin{pmatrix}
-\dfrac{3}{2}y^2 & -3xy+\sqrt{6} \epsilon y\\
-\sqrt{\dfrac{3}{2}}\lambda y\ \  &\ \dfrac{3}{2}-\dfrac{9}{2}y^2-\sqrt{\dfrac{3}{2}} \epsilon x
\end{pmatrix}.
\end{equation}
The matrix $\mathcal{M}$ has two two eigenvalues
for each critical points. 
The first critical point $\mathbf{X}_c^{(1)}$ is not stable, because the first eigenvalue is not negative, \textit{i.e.},
\begin{equation}
\mu_1^{(1)}=0\,,\quad \mu_2^{(1)}=\frac{3}{2}-\sqrt{\frac{3}{2}}\epsilon x\,.
\end{equation}
Concerning the second critical point $\mathbf{X}_c^{(2)}$, the eigenvalues
\begin{equation}
\mu_1^{(2)}=-\frac{3}{2}+\epsilon^2\,,\quad \mu_2^{(2)}=-3+2\epsilon^2
\end{equation}
are both negative in the existence domain $0<\epsilon\leq1$, implying that the critical point is stable and represents an attractor solution for the system (see \Cref{fig:background attractor}). 
Remarkably, for such a stable point, the inflation does not stop at arbitrary values of the slow-roll parameters, but it stops exactly once the conditions $\epsilon=|\eta|=1$ are attained.

\subsection{Chaotic behavior of the model}
\label{sec:chaotic}

\begin{figure*}
\begin{center}
\includegraphics[width=0.98\linewidth]{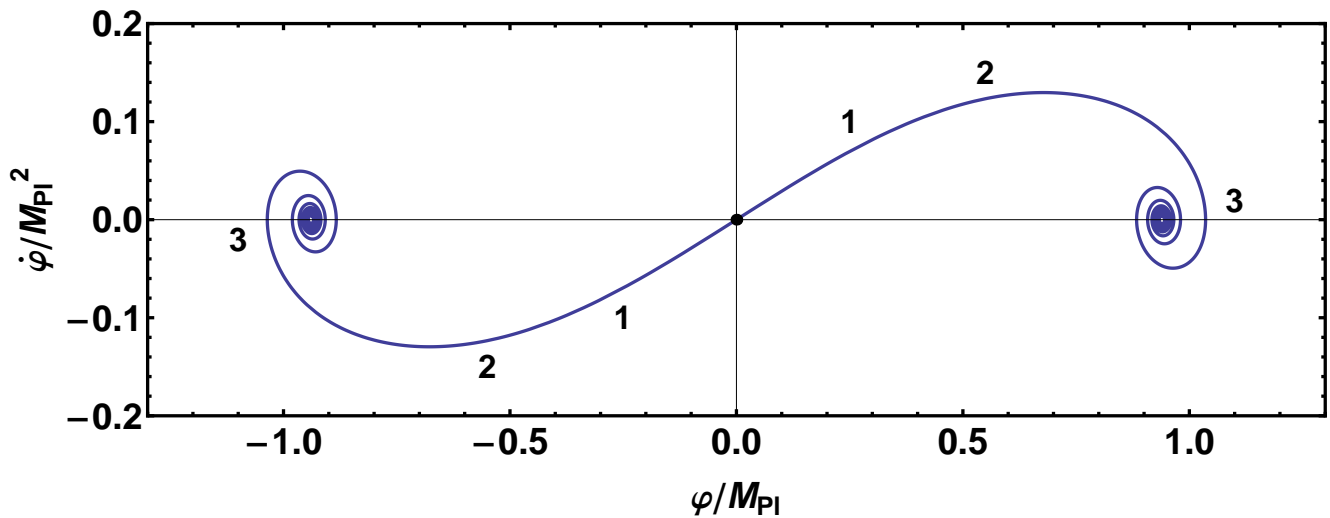}
\caption{Chaotic evolution of the system from $(0,0)$ (black dot) toward the minima $(\pm\varphi_0,0)$. The choice of the  minimum determines a spontaneous breaking of the symmetry, whereas the subsequent evolution exhibits a chaotic behavior. The physical features corresponding to regimes $1$, $2$ and $3$ are discussed in \Cref{sec:chaotic}.}
\label{fig:chaotic}
\end{center}
\end{figure*}

Here, we analyze the dynamics portrayed in \Cref{phieqofcont} without any approximation and fixing $\mathcal L_{,X}=1$.
In so doing, we write $\ddot \varphi=\dot\varphi_{,\varphi} \dot\varphi$, so that the equation of motion of the field $\varphi$ becomes a first-order differential equation 
\begin{equation}
\label{eq:chaotic}
\dot\varphi_{,\varphi}\dot{\varphi}+\frac{\sqrt{3\kappa^2}}{2}\sqrt{\dot\varphi^2+\mathcal V(\varphi)}\dot{\varphi}+\frac{\mathcal V_{,\varphi}(\varphi)}{2}=0\,.
\end{equation}
 The above relation is general and does not hold in the inflationary phase only. It consists in rewriting the first Friedmann equation, namely \Cref{1FriedSlow}, with our inflationary potential, considering no approximations over \Cref{phieqofcont}.
In this respect, the conditions $\mathcal L_{,X}\dot{\varphi}^2 \ll \mathcal V(\varphi)$ and $\ddot{\varphi} \ll 3H\dot{\varphi}/2$ are valid only during inflation, since they correspond to the slow roll hypothesis that only holds during phase $1$ of \Cref{fig:chaotic}. 

The generic solution of the above differential equation is shown in \Cref{fig:chaotic}, where we use $\varphi_0=0.94\,M_{\rm Pl}$, found above from the approximation $\varphi\ll1$.
The dynamics of the system starts from the point $(\varphi,\dot\varphi)=(0,0)$ and evolves toward the two minima $(\pm\varphi_0,0)$. Clearly, the system chooses one minimum, resulting in a spontaneous breaking of the symmetry. Moreover, the subsequent evolution toward the minimum exhibits a chaotic behavior.

We distinguish three different dynamical regimes:
\begin{enumerate}[1.]
    \item At the beginning, we have $\ddot\varphi\approx0$ and the potential $\mathcal V(\varphi)$ dominates over the kinetic term $\dot\varphi$. Thus, expanding around $\varphi\approx0$, from \Cref{2_3_5} we get
    \begin{equation}
        \dot\varphi=\frac{\varphi}{4}\sqrt{\frac{\kappa^2}{3}\mathcal V(0)} \propto\varphi\,,
    \end{equation}
    with $\mathcal V(0)=\chi\varphi_0^4/4$, which explains the linear evolution portrayed in  \Cref{fig:chaotic}.
    \item Afterwards, we still have $\ddot\varphi\approx0$ with $\dot\varphi$ dominating over $\mathcal V(\varphi)$, hence \Cref{2_3_5} becomes
    \begin{equation}
        \sqrt{3\kappa^2}\dot\varphi^2+\mathcal V_{,\varphi}(\varphi)=0\,.
    \end{equation}
    Again, expanding $\mathcal V_{,\varphi}(\varphi)$ around $\varphi\approx0$, we get
    \begin{equation}
        \dot\varphi=\pm\left[\frac{\kappa^2\mathcal V(0)^2}{3}\right]^{1/4}|\varphi|^{1/2}\propto|\varphi|^{1/2}\,,
    \end{equation}
    which explains the flattening in the evolution of $\dot\varphi$, in correspondence of the absolute maximum and minimum displayed in \Cref{fig:chaotic}.
    \item Finally, the field $\varphi$ oscillates around the minima. Expanding $\mathcal V(\varphi)$ around $\varphi_0$, and defining $\Delta\varphi\equiv \varphi-\varphi_0$ and $m\equiv\sqrt{\chi}\varphi_0 \ln 2$, from \Cref{1FriedSlow} we obtain the quadratic equation of an harmonic oscillator
    \begin{equation}
        \Delta\dot\varphi^2 + m^2\Delta\varphi^2 = \frac{3H^2}{\kappa^2}\,.
    \end{equation}
    Thus, using $\Delta\dot\varphi=\sqrt{3H^2/\kappa^2}\sin\theta$ and $m\Delta\varphi=\sqrt{3H^2/\kappa^2}\cos\theta$, where $\theta$ is the angular variable of the oscillations, and plugging into \Cref{phieqofcont}, after some cumbersome algebra we obtain
    \begin{equation}
        \label{oscillations}
        \dot\theta = - m - \frac{3}{4}H\sin(2\theta)\,,
    \end{equation}
    which explains the spiraling regime in \Cref{fig:chaotic}.
    Since $\dot H=-(3/2)H^2\sin^2\theta$, $H$ decreases, and so does the second term on the right side of \Cref{oscillations}, i.e. the oscillation amplitude.
\end{enumerate}

%%%%%%%%%%%%%%%%%%%%%%%%%%%%%%%%%%%%%%%%%%%%%%%%%%%%%%%%%%%%%%%%%%%%%%%%%%%%%%%%%%%%%%%%%%%%%%%%%%%%%%%%%%%%%%%%%%%%%%%%%%%%%%%%%%%%%%%%%%%%%%%%%%%%%%%%%%%%%%%%%%

\section{Theoretical discussion}\label{sezione7}

\begin{table*}
\footnotesize
\setlength{\tabcolsep}{0.5em}
\renewcommand{\arraystretch}{2}
\begin{tabular}{lcccccccccc}
\hline
\hline
$\qquad$ &  Before the transition    &  During the transition    &  After the transition    \\

\hline
Density       &     $\rho=2X\mathcal L_{,X}-K_0$
                & $\rho = 2X \mathcal{L}_{,X} + \mathcal V(\varphi)$ 
                & $
    \rho=2X\mathcal L_{,X}-K_0-\chi\varphi_0^4/4$
                \\
Pressure           & $P=K_0 $ 
                & $P=-\mathcal V(\varphi)$ 
                & $    P=K_0+\chi\varphi_0^4/4 $ 
                \\
Potential 
                & $V(0)=0$
                & $V(\varphi) = V_0+\frac{\chi\varphi_0^4}{4} \left[\frac{1-e^{-|\alpha|(\varphi^2-\varphi_0^2)}}{1-e^{|\alpha|\varphi_0^2}}\right]^2$ & $V(\varphi_0) = - \chi\varphi_0^4/4$ 
                \\
Prediction 
                &   $1^{st}$ order phase transition 
                &  Chaotic inflation 
                &  Dark energy
                \\
\hline
\hline
\end{tabular}
\caption{Schematic summary of the QQ model predictions. The cosmological constant problem is solved during the transition and predicts an inflationary phase, where the potential is built up from prime principles. The subsequent dynamics is accelerated after the transition. }
\label{Comparison1}
\end{table*}

Our paradigm aims to describe inflation during the phase transition induced by a symmetry-breaking potential (see \Cref{Comparison1}). In this scheme, the cosmological constant problem is greatly healed, as the degrees of freedom of vacuum energy are removed due to the cancellation mechanism of the proposed QQ model (see \Cref{Comparison2}). 
The corresponding inflationary potential has been reviewed in view of our current knowledge.\footnote{The initial scenario of inflation  has been extensively reviewed over the years. A large variety of models have been explored, such as the $R^2$ \cite{staro}, chaotic \cite{Chaotic}, power-law \cite{Lucchin85}, hybrid \cite{hybrid}, natural \cite{natural}, supernatural \cite{Randall96}, extranatural \cite{Arkani03}, eternal \cite{Guth07}, D-term \cite{Binetruy96}, F-term \cite{Casas99}, brane \cite{Dvali99}, oscillating \cite{Damour98}, trace anomaly driven \cite{Hawking01} etc.}
In particular, it is a \emph{type II inflation}, namely it provides ``small fields" since the inflaton field is small initially and slowly evolves toward the potential
minimum at larger $\varphi$. 
This scenario is contrasting with the rough classification made in \cite{Kolbclass}, where \emph{type I inflation} and \emph{type III inflation} have been also proposed.
In the first case, the initial  inflaton value is large and subsequently  rolls down toward the potential minimum at smaller values for $\varphi$, whereas in the second picture, the inflaton  is offered by hybrid (double) inflation model and inflation typically ends by the phase transition triggered by a second scalar field \cite{hybrid}.

As stressed above, our model is chaotic but, differently from the original proposal \cite{Chaotic}, that was a type I inflation, is quite similar to the new inflation \cite{Linde82,AS82} and natural
inflation \cite{natural}, which represent examples of type II inflation. Analogously to the latter cases, we here have $V_{,\varphi\varphi}(\varphi)>0$, but no sign changes as for natural inflation. 
We highly differ from type III inflation because there is not a second phase of inflation after the initial phase transition, arguable to heal the cosmological constant problem. 
In fact, our paradigm is quite similar to the Starobinsky inflationary scenario \cite{staro}, and the emerging potential interestingly recasts the Morse potential utilized in solid state physics \cite{Morse}.

\subsection{The coincidence and fine-tuning problems revised}

\begin{table*}
\footnotesize
\setlength{\tabcolsep}{0.5em}
\renewcommand{\arraystretch}{2}
\begin{tabular}{lcccccccccc}
\hline
\hline
Model &  Origin of $\Lambda$          &  Coincidence problem    
                &  Fine-tuning & Pressure & Density                 \\
\hline
$\Lambda$CDM       &     Quantum fluctuations
                & Yes 
                & Yes
                & Constant 
                & Constant
                \\
Dark fluid           & Quantum fluctuations 
                & Yes 
                & Yes
                & Constant
                & Variable
                \\
QQ 
                & Bare contribution 
                & No 
                & No
                & Constant
                & Variable\\

\hline
\hline
\end{tabular}
\caption{Physical differences among the $\Lambda$CDM model, the dark fluid approach and our QQ model applied to describe the universe dynamics.}
\label{Comparison2}
\end{table*}

The proposed QQ model revises the standard cosmological model and so both fine-tuning and coincidence issues might be healed. Here, we qualitatively debate a quantitative argument to see how to overcome the coincidence problem within our framework, while we underline that the fine-tuning issue is naturally fixed by means of the cancellation mechanism itself. 

Let us consider the standard Hilbert-Einstein Lagrangian density\footnote{Up to the volume of the universe, namely $ V\propto a(t)^{-3}$.} $\mathcal L = M_\text{Pl}^2R/2-\rho_{tot}$, where the Ricci scalar is given by $R=6(\dot H+2H^2)$, and $\rho_{tot}$ is the total density of the universe, regardless its constituents. Hence, after the transition $\rho_{tot}$ should be comparable to $R$. In fact, the coincidence problem can be easily addressed if, at the present time, $\left|P/\rho_{tot}\right|\sim 1$ \cite{coincidenza}. This suggests that, after the cancellation mechanism, it is quite unlikely that all vacuum energy is removed. This point is easily justifiable as, if one \emph{exactly} removes all vacuum energy, then a strong fine-tuning would be imposed by hand and so it is highly probable that fractions of critical density can therefore contribute to the net pressure.

Therefore, under such hypothesis, our dark fluid equation of state, for $\mathcal L_{,X}=1$, is given by\footnote{That fully degenerates with the total equation of state of $\Lambda$CDM.}
\begin{equation}\label{darkfluid}
\omega=-\frac{1}{1-\frac{2X}{K_0+\chi\varphi_0^4/4}}\,,
\end{equation}
which, compared to the total equation of state of the $\Lambda$CDM model, implies $X\equiv(1+z)^3$ and

\begin{equation}
    2\Omega_\Lambda=-\left(K_0+\frac{\chi\varphi_0^4}{4}\right)\Omega_m.
\end{equation} 
Now, it is evident that the case  $|K_0+\chi\varphi_0^4/4|\gg |2X|$ cannot occur because it happens before transition only. The case $|K_0+\chi\varphi_0^4/4|\ll |2X|$ cannot occur because it happens for baryons only, implying $\omega\rightarrow0$, and, by construction, $\varphi$ does not account for pressureless baryons. Hence, the only occurrence remains $|2X|\sim |K_0+\chi\varphi_0^4/4|$. However, since fractions of critical density may remain, after the transition, we clearly need that $|P|\sim |\rho_{tot}|\sim\rho_{cr}$,  healing \emph{de facto} the coincidence problem.

The fine-tuning issue does not occur at all, since during the transition vacuum energy disappears by virtue of the QQ cancellation mechanism. %The only question remained is how to transform the vacuum energy density into new species. In particular, a plausible scenario is that dark matter forms during the transition, implying new constituents with non-vanishing pressure. However, this leaves open the nature of such  constituents. Recent speculations have been carried out toward the production of geometric particles of dark matter that arise during this phase \cite{marca}. 
Finally, the physical features of our model compared to the standard cosmological model and to the dark fluid are summarized in \Cref{Comparison2}.

%%%%%%%%%%%%%%%%%%%%%%%%%%%%%%%%%%%%%%%%%%%%%%%%%%%%%%%%%%%%%%%%%%%%%%%%%%%%%%%%%%%%%%%%%%%%%%%%%%%%%%%%%%%%%%%%%%%%%%%%%%%%%%%%%%%%%%%%%%%%%%%%%%%%%%%%%%%%%%%%%%

\section{Outlooks and perspectives}\label{sezione8}

In this paper, we proposed that a QQ model, based on a new reformulation of the energy-momentum tensor for a scalar field, is able to solve the cosmological constant problem and to predict both accelerated phases of the universe, namely inflation and dark energy. To this aim, we first introduced a generic QQ model and then we showed how a Lagrangian for (dark) matter with pressure can reduce to the QQ landscape, under specific thermodynamic assumptions. We thus split the primordial universe into three main regimes, corresponding to pre-, during- and post-transition phases induced by a symmetry-breaking potential. Before the transition, the proposed potential is indistinguishable from the widely adopted fourth-order symmetry breaking potential. During the transition, under particular thermodynamic conditions, we obtained a Starobinsky-like potential, quite similar to solid state physics Morse potential. Our inferred potential  breaks the symmetry and suggests the existence of an inflationary phase from small to large fields. In particular, our inflationary scenario  unifies the pictures provided by old and new inflation  into a single scheme. To clarify this fact, we underlined that in the standard cosmological puzzle either phase transition or chaotic inflation can explain the inflationary phase. On the contrary, in our scheme, the phase transition induces the mechanism responsible for removing vacuum energy, whereas chaotic inflation is responsible for the graceful exit from inflation. Both landscapes coexist into our inflationary potential, by simply requiring to cancel out the vacuum energy quantum fluctuations.

We then investigated our inflationary potential and discussed the critical points and their stability. Moreover, we showed in detail how to heal the cosmological constant problem during the transition and how to pass to the third phase. The third phase, namely after the transition, has been here reinterpreted, showing that dark energy dynamics can be recovered as the universe expands. We demonstrated under which conditions our QQ model resembled the dark fluid behavior, mimicking the $\Lambda$CDM framework. In this respect, the proposed paradigm extends the mechanism proposed in \cite{Luongo18}, unifying inflation with dark energy. In fact,  if one solves the cosmological constant problem, from our paradigm it is then possible to recover the functional behavior of the standard cosmological model, having however a bare cosmological constant that contributes to dark energy, i.e., without any coincidence and fine-tuning issues.  

Future works will involve the study of more complicated potentials during the transition, and the inclusion within this scheme of baryogenesis before transition. We will also investigate whether our QQ dark fluid could be able to solve the current tensions on cosmological parameters.

%%%%%%%%%%%%%%%%%%%%%%%%%%%%%%%%%%%%%%%%%%%%%%%%%%%%%%%%%%%%%%%%%%%%%%%%%%%%%%%%%%%%%%%%%%%%%%%%%%%%%%%%%%%%%%%%%%%%%%%%%%%%%%%%%%%%%%%%%%%%%%%%%%%%%%%%%%%%%%%%%%

\acknowledgments 
R.D. acknowledges the support of INFN (\emph{iniziativa specifica} QGSKY). O.L. and M.M. acknowledge the support of the Ministry of Education and Science of the Republic of Kazakhstan, Grant IRN AP08052311. \\

%%%%%%%%%%%%%%%%%%%%%%%%%%%%%%%%%%%%%%%%%%%%%%%%%%%%%%%%%%%%%%%%%%%%%%%%%%%%%%%%%%%%%%%%%%%%%%%%%%%%%%%%%%%%%%%%%%%%%%%%%%%%%%%%%%%%%%%%%%%%%%%%%%%%%%%%%%%%%%%%%%

\end{document}